\documentclass[aps,nofootinbib,prd,superscriptaddress,tightenlines,notitlepage,
twocolumn,showpacs,floatfix]{revtex4-1}
\usepackage[colorlinks,citecolor=blue]{hyperref}
\usepackage{tabularx}
\usepackage{graphicx}
\usepackage{amssymb,bm,tensor}
\usepackage{xcolor}
\usepackage{amsmath}
\usepackage[varg]{txfonts}
\usepackage{enumerate}
\usepackage[normalem]{ulem}
\usepackage{bm}% bold math
\usepackage[OT1]{fontenc}
\usepackage{orcidlink}

\begin{document}

\title{Galactic double neutron stars as dual-line gravitational-wave sources: \\ 
Prospects with LISA and Cosmic Explorer}

\author{Wen-Fan Feng\orcidlink{0000-0001-5033-6168}}
\email{fengwf@pku.edu.cn}
\affiliation{Kavli Institute for Astronomy and Astrophysics, Peking University,
Beijing 100871, China}

\author{Lijing Shao\orcidlink{0000-0002-1334-8853}}
\email{lshao@pku.edu.cn}
\affiliation{Kavli Institute for Astronomy and Astrophysics, Peking University,
Beijing 100871, China}%
\affiliation{National Astronomical Observatories, Chinese Academy of Sciences,
Beijing 100012, China}

\date{\today}

\begin{abstract}

Double neutron star (DNS) systems could serve as intriguing dual-line
gravitational-wave (GW) sources, emitting both high- and low-frequency GWs,
arising respectively from the asymmetric spinning bodies of individual neutron
stars (NSs) and the binary orbital inspiral. Detecting such dual-line signals
could provide novel perspectives on binary orbital geometry and NS internal
physics. We expand upon previously calculated spinning NS waveforms by
incorporating higher-order terms of NS structural parameters. A population
simulation is performed for spinning NS components in DNS systems potentially
detectable by the space-based Laser Interferometer Space Antenna (LISA). Based
on 4-year LISA observation of 35 resolvable DNS systems under an optimistic scenario, 
we estimate that 6 (22) spinning NS components could be detected by the next-generation ground-based GW
detector, Cosmic Explorer, under log-uniform (uniform) sampling of NS structural
parameters. For these dual-line sources, the median signal-to-noise ratio
achievable with Cosmic Explorer is approximately 20--30. Through the dual-line
GW detection strategy, the relative measurement accuracy of the NS moment of
inertia is estimated to be $\sim 8\%$.

\end{abstract}

\maketitle

\section{Introduction}

The era of gravitational-wave (GW) astronomy has opened a novel observational
frontier for exploring the universe \cite{LIGOScientific:2016aoc,
LIGOScientific:2017vwq, KAGRA:2021vkt}. While Advanced LIGO
\cite{advancedLIGO2015} and Advanced Virgo \cite{AdvancedVirgo2015} have
detected burst signals from compact binary mergers, the broader GW spectrum
motivates intensive studies of diverse signal types, enabled by next-generation
detectors designed for complementary frequency regimes \cite{KAGRA:2013rdx,
Kalogera:2021bya, Abac:2025saz, ZhaoWen:2018ztu, Kuroda:2015owv, Shao:2020}.

Double neutron stars (DNSs) are prime GW sources due to the unique dual-line
emissions: hectohertz high-frequency GWs arise from the intrinsic rotational
asymmetry of individual neutron stars (NSs), while millihertz low-frequency GWs
emerge from binary orbital inspiral. Upcoming ground-based detectors---Cosmic
Explorer \cite{CE2022slt} and Einstein Telescope \cite{ET2010}---operating in
tandem with space-borne observatories like LISA \cite{LISA2017}, TianQin
\cite{TianQin2016}, and Taiji \cite{Taiji2017}---are poised to detect these
dual-line signals. Such multiband observations will deepen our understanding of
the fundamental physical theories for binary orbital dynamics and NS internal
structure \cite{Feng:2023gpe, Feng:2024ulg}.

Previous studies of dual-line GW sources have predominantly focused on
ultracompact X-ray binaries (UCXBs), with a specific emphasis on NS-white dwarf
(WD) systems \cite{Tauris:2018kzq, Chen:2021mrf, Suvorov:2021mhr}.  Binary
stellar evolution simulations have shown that low-frequency GWs from a system
comprising a $1.6~M_{\odot}$ NS and a $0.16~M_{\odot}$ helium WD can enter the
LISA sensitivity band \cite{Tauris:2018kzq}. Owing to the narrow mass range of
the helium WD, NS mass measurements in such systems achieve percent-level
precision. By analyzing the amplitudes of dual-line strain signals, constraints
can be placed on the combination of the NS moment of inertia and equatorial
ellipticity \cite{Tauris:2018kzq}.  Further, assuming that the NS spin-down is
dominated by gravitational radiation rather than electromagnetic (EM) processes,
pulsar timing observations of binary pulsars might enable separate inference of
the NS moment of inertia and equatorial ellipticity \cite{Chen:2021mrf}.  From a
detection perspective, the dual-line detectability of 12 known UCXBs with
sub-hour orbital periods was assessed, considering GW strains from different
radiation mechanisms \cite{Suvorov:2021mhr}.  For dual-line DNS systems in the
millihertz band, \citet{Feng:2023gpe} demonstrated that joint dual-line
detection---using waveforms incorporating spin-orbit coupling---can constrain
the binary orbital geometry and the moment of inertia of NSs.

Over the past two decades, numerous studies have explored the detectability of
continuous GWs from Galactic isolated NSs via population synthesis simulation
\cite{Palomba:2005na, Knispel:2008ue, Wade:2012qc, Cieslar:2021viw,
Reed:2021scb, Soldateschi:2021hfk, Pagliaro:2023bvi, Hua:2023aff}, known pulsars
\cite{Pitkin:2011sc, LIGOScientific:2017hal, LIGOScientific:2019xqs,
LIGOScientific:2020gml, Beniwal:2021hvc, LIGOScientific:2021hvc,
LIGOScientific:2025kei}, and unknown NSs in long-period binary systems
neglecting their interactions with the companion stars \cite{Leaci:2015bka,
Covas:2019jqa, LIGOScientific:2020qhb, Covas:2020nwy, Zhang:2020rph,
Covas:2024nzs}.  Population synthesis of isolated (GW-driven or EM-driven) NSs
usually requires specifying the distributions for spatial locations, kick
velocities, initial spin periods, ages, and equatorial ellipticity parameters.
The number of detectable Galactic DNSs in LISA or TianQin has been predicted
using both binary population synthesis \cite{Nelemans:2001hp, Belczynski:2008nh,
Liu:2014qaa, Lau:2019wzw, Breivik:2019lmt, Wagg:2021cst} and simplified backward
evolution simulations \cite{Andrews:2019plw, Feng:2023fez}.

However, no prior simulation has yet integrated the modeling of the DNS
population with that of the individual spinning NS component population.  In
this work, we first expand upon previously calculated GWs emitted by triaxially
deformed precessing NSs under spin-orbit coupling effects \cite{Feng:2024ect},
by assuming that the nonsphericity and wobble angle of NSs are independent
\cite{Gao:2020zcd}. As a preliminary exploration of dual-line GW sources, we
assign spatial positions and orbital parameters to the spinning NS components
based on LISA-detectable DNS systems from COMPAS population synthesis
simulations \cite{Wagg:2021cst, COMPASTeam:2021tbl}. By performing uniform and
log-uniform sampling within the predefined ranges of NS characteristic
parameters, we generate two corresponding populations of spinning NS components.
We analyze the distributions of dual-line detectability counts and
signal-to-noise ratios (SNRs) for these populations with the Cosmic Explorer
detector. Finally, we investigate the scientific implications of detecting
dual-line GW sources, focusing specifically on their role in measuring the
moment of inertia of NSs.

The paper is structured as follows. In Sec.~\ref{sec-PopulationSynthesis}, we
characterize the parameter distributions of the LISA-detectable DNS systems via
binary population synthesis, derive GW waveforms of NSs in binary systems
incorporating high-order expansions of characteristic parameters, and perform
population synthesis of spinning NS components within these systems. In
Sec.~\ref{sec-results}, we combine LISA and Cosmic Explorer observations to
estimate the detectable numbers and SNRs of dual-line DNS sources and infer the
moment of inertia distribution for NSs. Our conclusions are presented in
Sec.~\ref{sec-conclusion}.

\section{Population synthesis simulations of dual-line GW sources}
\label{sec-PopulationSynthesis}

\subsection{Parameter distributions of LISA-detectable DNSs via population synthesis}
\label{sec-DNSPopulationSynthesis}

Future space-based GW detectors like LISA will be sensitive to compact binary
systems in the Milky Way. Using an empirically informed Milky Way model of the
metallicity-dependent star formation history and extensive binary population
synthesis predictions with 20 model variations (accounting for uncertainties in
binary evolution)\cite{Broekgaarden:2021iew, Broekgaarden:2021efa}, \citet{Wagg:2021cst} estimates their detectability. For a
4-year LISA mission, the expected number of DNS detections is 3--35. The
parameter distributions of 35 LISA-detectable DNS systems are summarized below
\cite{Wagg:2021cst}, forming the basis for our study of the NS population in
Sec.~\ref{sec-NSPopulationSynthesis}.

\begin{itemize}
    \item {\bf Component mass}: Most DNSs exhibit a mass ratio close to unity,
    with $90\%$ systems having a ratio $>0.8$. This tendency toward equal masses
    arises because the majority of NSs in the simulations either form through
    electron-capture supernovae (assuming a remnant mass of $1.26~M_{\odot}$) or
    originate from low-mass stars. The remnant mass prescription assumes a
    constant fallback mass for any star with a carbon-oxygen core mass below
    $2.5~M_{\odot}$, resulting in many NSs acquiring a mass of
    $1.278~M_{\odot}$.
    \item {\bf Spatial position}: Most detectable DNSs are concentrated near the
    Galactic center with a distance distribution peaking around 8 kpc, showing a
    significant bias toward sources on our side of the Milky Way, in the Solar
    system's vicinity. Their detectable range reaches
    $\sim 20~\rm{kpc}$, though such distant systems are rare (see also Ref.~\cite{Wagg:2021sgn} for horizon distance discussion). About $90\%$ of sources have right ascensions (RA)
    within $200^{\circ}$ to $300^{\circ}$, and declinations (DEC) within
    $-70^{\circ}$ to $30^{\circ}$.
    \item {\bf Orbital frequency}: The distribution is nearly symmetric, peaking
    at $0.6~{\rm mHz}$ (corresponding to an orbital period of 28 min).
    \item {\bf Orbital eccentricity}: The distribution peaks at 0.01, with about
    $90\%$ of sources having eccentricities $< 0.2$.
    \item {\bf Signal-to-noise ratio}: The SNR distribution is similar to that
    of orbital eccentricity. Most sources have SNRs concentrated between 8 and
    60, with about $90\%$ of the sources below 100. 
\end{itemize}

\subsection{GW modeling from spinning NS components in LISA-detectable DNSs}
\label{sec-NSGWmodeling}

When calculating the waveforms radiated by a spinning NS component in a typical
LISA-detectable binary, the spin-orbit coupling is an important factor
influencing the waveform modulation \cite{Feng:2023gpe}. In a simple precession
scheme, we focus on the spinning NS emitting continuous GWs, while neglecting the spin of its companion star. The precession under consideration here is the spin–orbit precession, commonly referred to as geodetic precession \cite{poisson2014gravity}. This precession modulation is determined by the average precession angular
frequency around the binary's total angular momentum $\boldsymbol{J}$ (with its
magnitude $J$)
\begin{equation}
\Omega_{\rm pre} = \frac{G J}{c^{2} a^{3}(1-e^2)^{3/2}}\left(2+\frac{3 m_2}{2 m_{1}}\right)  \,,
\end{equation}
where $m_{1}$ and $m_{2}$  are the NS component masses, $a$ and $e$ are the
semimajor axis and orbital eccentricity of the binary system.  For a typical DNS
system with a 10-min orbital period, the precession period induced by spin-orbit
coupling is $\mathcal{O}(1\,\rm{yr})$. The key quantity entering the waveforms
is the precession angle of the NS component's spin angular momentum, defined as
$\alpha = \Omega_{\rm pre} t$ \cite{Feng:2024ect}, with the initial phase set to
zero.

Waveforms emitted by a precessing triaxial NS are conventionally described via a
series expansion in small parameters, as detailed in previous works
\cite{Zimmermann:1980ba, VanDenBroeck:2004wj, Gao:2020zcd, Gao:2022hzd,
Feng:2023gpe}. 
In addition to the common gravitational radiation at twice the spin frequency, precession introduces a non-zero contribution in the quadrupolar radiation at the fundamental spin frequency.
To streamline the theoretical framework, we define the angular
frequencies of free precession ($\Omega_{\rm p}$) and rotation ($\Omega_{\rm
r}$) for the spinning NS, alongside with three dimensionless small parameters
characterizing its structure: oblateness (or poloidal ellipticity) $\epsilon$,
nonsphericity $\kappa$, and wobble angle $\gamma$. These are explicitly given
by:
\begin{align}\label{eq_smallparas}
\epsilon \equiv \frac{I_3-I_1}{I_3} \,,\quad
\kappa   \equiv \frac{1}{16}\frac{I_3}{I_1}\frac{I_2-I_1}{I_3-I_2} \,,\quad
\gamma   \equiv \frac{\omega_{10} I_1}{\omega_{30} I_3} \,,
\end{align}
where $\omega_{10}$ and $\omega_{30}$ denote the initial spin angular velocity components for $\omega_1$ and $\omega_3$ in body frame of the NS, respectively, and $(I_1, I_2, I_3)$ represent the principal
moments of inertia with $I_3 > I_2 > I_1$.
The parameter $\Omega_{\rm p}$ characterizes the rate at which the spin angular velocity rotates about the three principal axes of inertia in body frame of the NS, while $\Omega_{\rm r}$ denotes the rate of rotation about the spin angular momentum vector in inertial frame. Detailed expressions for these parameters can be found in Eqs. (6)-(12) in \cite{Feng:2023gpe}. For a representative DNS system with a rapidly rotating NS component, $\Omega_{\rm r} \simeq 2\pi \times 100~{\rm Hz}$ and $\Omega_{\rm p} \simeq 2~{\rm mHz}$ (see the caption of Figure 3 in \cite{Feng:2023gpe}).

Compared with the hierarchical relation $\kappa \sim \mathcal{O}(\gamma^{2})$
adopted in previous studies (e.g., Refs.~\cite{Feng:2023gpe,
VanDenBroeck:2004wj}), we treat $\kappa$ and $\gamma$ as independent small
parameters \cite{Gao:2020zcd}. Accordingly, we extend the waveform expansion
from Ref.~\cite{Feng:2023gpe} to the order $\mathcal{O}(\gamma^2, \kappa^2)$,
explicitly incorporating contributions from the cross term
$\mathcal{O}(\gamma\kappa)$ and the quadratic term $\mathcal{O}(\kappa^{2})$.

We denote the waveform components with distinct frequencies emitted by the
spinning NS as $h_{+}^{(\mu x)}$ and $h_{\times}^{(\mu x)}$, where the index
$\mu = {1,2}$ labels emissions near the spin frequency and twice the spin
frequency, respectively, and $x = a,b,c$ corresponds to different GW
frequencies.  The two GW polarizations generated by the spinning NS undergoing
spin precession can be expanded in terms of the characteristic parameters
defined in Eq.~(\ref{eq_smallparas}) as follows:
\begin{subequations}  \label{eq_WaveformComponents}
\begin{align}
h_+ &= h_+^{(1a)} + h_+^{(1b)} + h_+^{(1c)} + h_+^{(2a)} + h_+^{(2b)} + h_+^{(2c)} + \cdots  \,, \\
h_\times &= h_\times^{(1a)} + h_\times^{(1b)} + h_\times^{(1c)} + h_\times^{(2a)} + h_\times^{(2b)} + h_\times^{(2c)} + \cdots \,,
\end{align}
\end{subequations}
where
\begin{widetext}
\begin{subequations} 
\begin{align} 
\nonumber
 h_{+}^{(1a)} = &\frac{G b^2 I_3 \epsilon (\gamma-8\gamma\kappa)}{4 c^4 D}\bigg\{ \Big[\sin 2 \theta_S \big(6 \sin ^2 \iota-(3+\cos 2 \iota) \cos 2 \alpha \big)+4 \cos 2 \theta_S \cos \alpha \sin 2 \iota \Big]  \cos \left[\left(\Omega_{\mathrm{r}}+\Omega_{\mathrm{p}}\right) t\right] \\ 
&  +2\Big[-2 \cos \theta_S \sin 2 \iota \sin \alpha+ \big(3+\cos 2 \iota \big) \sin \theta_S \sin 2 \alpha \Big] \sin \left[\left(\Omega_{\mathrm{r}}+\Omega_{\mathrm{p}}\right)t\right]\bigg\} , \\ \nonumber
h_{+}^{(1b)} = &\frac{5G b^2 I_3 \epsilon \gamma\kappa}{2 c^4 D}\bigg\{ \Big[\sin 2 \theta_S \big(6 \sin ^2 \iota-(3+\cos 2 \iota) \cos 2 \alpha \big)+4 \cos 2 \theta_S \cos \alpha \sin 2 \iota \Big]  \cos \left[\left(\Omega_{\mathrm{r}}-\Omega_{\mathrm{p}}\right) t\right] \\ 
&  +2\Big[-2 \cos \theta_S \sin 2 \iota \sin \alpha+ \big(3+\cos 2 \iota \big) \sin \theta_S \sin 2 \alpha \Big] \sin \left[\left(\Omega_{\mathrm{r}}-\Omega_{\mathrm{p}}\right)t\right]\bigg\} , \\ \nonumber
h_{+}^{(1c)} = & -\frac{G b^2 I_3 \epsilon \gamma\kappa}{2 c^4 D}\bigg\{ \Big[\sin 2 \theta_S \big(6 \sin ^2 \iota-(3+\cos 2 \iota) \cos 2 \alpha \big)+4 \cos 2 \theta_S \cos \alpha \sin 2 \iota \Big]  \cos \left[\left(\Omega_{\mathrm{r}} + 3\Omega_{\mathrm{p}}\right) t\right] \\ 
&  +2\Big[-2 \cos \theta_S \sin 2 \iota \sin \alpha+ \big(3+\cos 2 \iota \big) \sin \theta_S \sin 2 \alpha \Big] \sin \left[\left(\Omega_{\mathrm{r}} + 3\Omega_{\mathrm{p}}\right)t\right]\bigg\} , \\ \nonumber
 h_{+}^{(2a)} = &-\frac{16 G b^2 I_3 \epsilon (\kappa-16\kappa^2)}{c^4 D}\left[(3+\cos (2 \iota))\left[\cos ^4\left(\frac{\theta_S}{2}\right) \cos \left(2 \alpha+2t\Omega_{\mathrm{r}}\right)+\cos \left(2\alpha-2t\Omega_{\mathrm{r}}\right) \sin ^4\left(\frac{\theta_S}{2}\right)\right]\right. \\ 
& \left.+\cos \left(2 t \Omega_{\mathrm{r}}\right)\left[3 \sin ^2 \theta_S \sin ^2 \iota+\cos \alpha \sin \left(2 \theta_S\right) \sin (2 \iota)\right]-2 \sin \theta_S \sin (2 \iota) \sin \alpha \sin \left(2 t \Omega_{\mathrm{r}}\right)\right], \\ \nonumber
 h_{+}^{(2b)} = &\frac{G b^2 I_3 \epsilon (\gamma^2+64\kappa^2)}{c^4 D}\left[(3+\cos (2 \iota))\left[\cos ^4\left(\frac{\theta_S}{2}\right) \cos \left(2\alpha+2 t\left(\Omega_{\mathrm{r}}+\Omega_{\mathrm{p}}\right)\right)+\cos \left(2\alpha-2 t\left(\Omega_{\mathrm{r}}+\Omega_{\mathrm{p}}\right)\right) \sin ^4\left(\frac{\theta_S}{2}\right)\right]\right. \\ 
& \left.+\cos \left(2 t\left(\Omega_{\mathrm{r}}+\Omega_{\mathrm{p}}\right)\right)\left[3 \sin ^2 \theta_S \sin ^2 \iota+\cos \alpha \sin \left(2 \theta_S\right) \sin (2 \iota)\right]-2 \sin \theta_S \sin (2 \iota) \sin \alpha \sin \left(2 t\left(\Omega_{\mathrm{r}}+\Omega_{\mathrm{p}}\right)\right)\right], \\ \nonumber
 h_{+}^{(2c)} = &-\frac{64 G b^2 I_3 \epsilon \kappa^2}{c^4 D}\left[(3+\cos (2 \iota))\left[\cos ^4\left(\frac{\theta_S}{2}\right) \cos \left(2\alpha+2 t\left(\Omega_{\mathrm{r}}-\Omega_{\mathrm{p}}\right)\right)+\cos \left(2\alpha-2 t\left(\Omega_{\mathrm{r}}-\Omega_{\mathrm{p}}\right)\right) \sin ^4\left(\frac{\theta_S}{2}\right)\right]\right. \\ 
& \left.+\cos \left(2 t\left(\Omega_{\mathrm{r}}-\Omega_{\mathrm{p}}\right)\right)\left[3 \sin ^2 \theta_S \sin ^2 \iota+\cos \alpha \sin \left(2 \theta_S\right) \sin (2 \iota)\right]-2 \sin \theta_S \sin (2 \iota) \sin \alpha \sin \left(2 t\left(\Omega_{\mathrm{r}}-\Omega_{\mathrm{p}}\right)\right)\right]  \,,
\end{align}
\end{subequations}
\end{widetext}
and
\begin{widetext}
\begin{subequations} 
\begin{align} 
\nonumber
h_{\times}^{(1a)} = &\frac{G b^2 I_3 \epsilon (\gamma-8\gamma\kappa)}{c^4 D}\left[\cos \left(t\left(\Omega_{\mathrm{r}}+\Omega_{\mathrm{p}}\right)\right)\left[2 \sin \iota \cos \left(2 \theta_S\right) \sin \alpha-\cos \iota \sin \left(2 \theta_S\right) \sin \left(2 \alpha\right)\right]\right. \\ 
& \left.+2\left[\sin \iota \cos \theta_S \cos \alpha-\cos \iota \cos \left(2 \alpha\right) \sin \theta_S\right] \sin \left(t\left(\Omega_{\mathrm{r}}+\Omega_{\mathrm{p}}\right)\right)\right], \\ \nonumber
h_{\times}^{(1b)} = &\frac{10 G b^2 I_3 \epsilon \gamma \kappa}{c^4 D}\left[\cos \left(t\left(\Omega_{\mathrm{r}}-\Omega_{\mathrm{p}}\right)\right)\left[2 \sin \iota \cos \left(2 \theta_S\right) \sin \alpha-\cos \iota \sin \left(2 \theta_S\right) \sin \left(2 \alpha\right)\right]\right. \\ 
& \left.+2\left[\sin \iota \cos \theta_S \cos \alpha-\cos \iota \cos \left(2 \alpha\right) \sin \theta_S\right] \sin \left(t\left(\Omega_{\mathrm{r}}-\Omega_{\mathrm{p}}\right)\right)\right], \\ \nonumber
h_{\times}^{(1c)} = &-\frac{2G b^2 I_3 \epsilon \gamma\kappa}{c^4 D}\left[\cos \left(t\left(\Omega_{\mathrm{r}}+3\Omega_{\mathrm{p}}\right)\right)\left[2 \sin \iota \cos \left(2 \theta_S\right) \sin \alpha-\cos \iota \sin \left(2 \theta_S\right) \sin \left(2 \alpha\right)\right]\right. \\ 
& \left.+2\left[\sin \iota \cos \theta_S \cos \alpha-\cos \iota \cos \left(2 \alpha\right) \sin \theta_S\right] \sin \left(t\left(\Omega_{\mathrm{r}}+3\Omega_{\mathrm{p}}\right)\right)\right], \\ \nonumber
h_{\times}^{(2a)} = &\frac{16 G b^2 I_3 \epsilon (\kappa-16\kappa^2)}{c^4 D}\left[\cos \left(2 t \Omega_{\mathrm{r}}\right)\left[-2 \sin \iota \sin \left(2 \theta_S\right) \sin \alpha-\cos \iota\left(3+\cos \left(2 \theta_S\right)\right) \sin \left(2 \alpha\right)\right]\right. \\ 
& \left.-4\left[\cos \iota \cos \theta_S \cos \left(2 \alpha\right)+\sin \iota \cos \alpha \sin \theta_S\right] \sin \left(2 t \Omega_{\mathrm{r}}\right)\right], \\ \nonumber
 h_{\times}^{(2b)} = &\frac{G b^2 I_3 \epsilon  (\gamma^2+64\kappa^2)}{c^4 D}\left[\cos \left(2 t\left(\Omega_{\mathrm{r}}+\Omega_{\mathrm{p}}\right)\right)\left[2 \sin \left(2 \theta_S\right) \sin \iota \sin \alpha+\left(3+\cos \left(2 \theta_S\right)\right) \cos \iota \sin \left(2 \alpha\right)\right]\right. \\ 
& \left.+4\left[\cos \theta_S \cos \left(2 \alpha\right) \cos \iota+\cos \alpha \sin \theta_S \sin \iota\right] \sin \left(2 t\left(\Omega_{\mathrm{r}}+\Omega_{\mathrm{p}}\right)\right)\right] , \\ \nonumber
 h_{\times}^{(2c)} = &-\frac{64 G b^2 I_3 \epsilon \kappa^2}{c^4 D}\left[\cos \left(2 t\left(\Omega_{\mathrm{r}}-\Omega_{\mathrm{p}}\right)\right)\left[2 \sin \left(2 \theta_S\right) \sin \iota \sin \alpha+\left(3+\cos \left(2 \theta_S\right)\right) \cos \iota \sin \left(2 \alpha\right)\right]\right. \\ 
& \left.+4\left[\cos \theta_S \cos \left(2 \alpha\right) \cos \iota+\cos \alpha \sin \theta_S \sin \iota\right] \sin \left(2 t\left(\Omega_{\mathrm{r}}-\Omega_{\mathrm{p}}\right)\right)\right] .
\end{align}
\end{subequations}
\end{widetext}
These waveforms depend on the binary inclination angle $\iota$, spin precession cone opening angle $\theta_S$, and precession angle $\alpha$ \cite{Feng:2024ect, Feng:2023gpe}.
In the limit as $\Omega_{\rm pre} \to 0$, the plus polarizations are
\begin{subequations}
\begin{align}
h_{+}^{(1a)}  \propto & \, \mathcal{O}(\gamma) \cos [(\Omega_{\rm r}+\Omega_{\rm p})t]  \,, \\  
h_{+}^{(1b)}  \propto & \, \mathcal{O}(\gamma\kappa) \cos [(\Omega_{\rm r}-\Omega_{\rm p})t]  \,, \\ 
h_{+}^{(1c)}  \propto & \, \mathcal{O}(\gamma\kappa) \cos [(\Omega_{\rm r}+3\Omega_{\rm p})t]  \,, \\
h_{+}^{(2a)}  \propto & \, \mathcal{O}(\kappa) \cos (2\Omega_{\rm r}t)  \,, \\  
h_{+}^{(2b)}  \propto & \, [\mathcal{O}(\gamma^2)+\mathcal{O}(\kappa^2)] \cos [2(\Omega_{\rm r}+\Omega_{\rm p})t]  \,, \\ 
h_{+}^{(2c)}  \propto & \, \mathcal{O}(\kappa^2) \cos [2(\Omega_{\rm r}-\Omega_{\rm p})t]  \,.
\end{align}
\end{subequations}
The components of $h_{\times}^{(\mu x)} (\Omega_{\rm pre} \to 0)$ are similar to
that of $h_{+}^{(\mu x)} (\Omega_{\rm pre} \to 0)$, with the cosine function in
the latter being replaced by the sine function. These can be reduced to the
expressions in Ref.~\cite{Gao:2020zcd}. 
% with the high-order terms $\mathcal{O}(\gamma\kappa)$ and
% $\mathcal{O}(\kappa^{2})$ in $h_{+}^{(1a)}$ and $ h_{+}^{(2a)}$ dropped.

\subsection{Parameters of spinning NS components in LISA-detectable DNSs}
\label{sec-NSPopulationParameters}

According to the calculation in Sec.~\ref{sec-NSGWmodeling}, the GWs emitted by
spinning NSs in tight binaries under spin-orbit coupling depend on the spin
period, NS characteristic parameters associated with the principal moment of
inertia, the spin precession cone opening angle, and the binary inclination
angle.  We describe these parameters in detail in the following subsections.

\subsubsection{spin period}\label{subsec-spin}

The majority of DNS systems in Ref.~\cite{Wagg:2021cst} form through
electron-capture supernovae and ultra-stripped supernovae. For these types of
supernovae, natal kick magnitudes follow a Maxwellian velocity distribution with
a one-dimensional root-mean-square (rms) velocity dispersion of $\sigma_{\rm
rms} = 30~\rm{km~s}^{-1}$. According to the population synthesis in
Ref.~\cite{Dewi:2005xh}, a correlation between pulsar spin period and orbital
eccentricity in Galactic DNSs emerges when the second NS receives a low kick
with velocity dispersion below $50~\rm{km~s}^{-1}$. Thus, we simulate the NS
spin period based on this correlation. 

% The gray data points were acquired by KDE to determine the most probable spin
% period $P$ values corresponding to specific orbital eccentricity $e$ values in
% DNSs. A grid of eccentricity values is created from 0 to 1 in 0.05 increments.
% The second term on the right-hand side of the above equation accounts for the
% dispersion of the $(e, P)$ data points by adding a normally distributed noise
% term with a mean of 0 and a standard deviation of $\sigma_P=3\sigma_{\rm fit}$
% to the linear relation mentioned above. The term $\sigma_{\rm fit}$ originates
% from the standard deviation of the fitting residuals, and amplifying it by 3
% times is intended to account for extreme fluctuations in the data
% conservatively. 

We derive this correlation through the following procedure.  We employ kernel
density estimation (KDE) to determine the most probable spin period ($P$) for a
given orbital eccentricity ($e$), facilitating the establishment of their
relation. First, we filter data points in Ref.~\cite{Dewi:2005xh} by selecting
those within a pre-defined bandwidth around a target $e$-value, where the
absolute difference between their eccentricities and the target defines the
relevant range. Next, we construct a kernel density distribution for the
filtered spin periods to estimate their probability density, identifying the
peak of the probability density function (PDF) via maximization around the mean
of the filtered values—this peak denotes the most probable $P$ for the given
$e$. A grid of $e$ values spanning from 0 to 1 in 0.05 increments is generated,
and for each $e$, the above process is repeated with a fixed bandwidth, yielding
paired $(P, e)$ values.  

The spin period–orbital eccentricity relation is ultimately fitted for the above
selected data points of $(P,e)$ as
\begin{equation}\label{eq-fittedline}
P/[{\rm ms}] \simeq 163~e+19.7 \,,
\end{equation}
with a coefficient of determination $R^2=0.95$ and the rms error $\sigma_{\rm
fit}=11.1~\rm{ms}$.  To characterize the dispersion in the simulated data of
$(P,e)$ distribution, a Gaussian noise term with zero mean and a standard
deviation of $\sigma_P = 3\sigma_{\rm fit}$ is added to the fitted linear model:
\begin{equation}\label{eq-fittednoise}
\delta P/[\rm{ms}] \sim \mathcal{N}(0,\sigma_P^2) \,.
\end{equation}
Using $\sigma_P =3\sigma_{\rm fit}$ as the noise standard deviation effectively reproduces
the dispersion of simulated data points. We also dynamically calculate the lower
limit for the noise to ensure that the simulated spin period satisfies
$P>10~{\rm ms}$, which is usually assumed for recycled NSs in Galactic DNSs.

\begin{figure}[t]%[!htbp]
\centering
\includegraphics[scale=0.6]{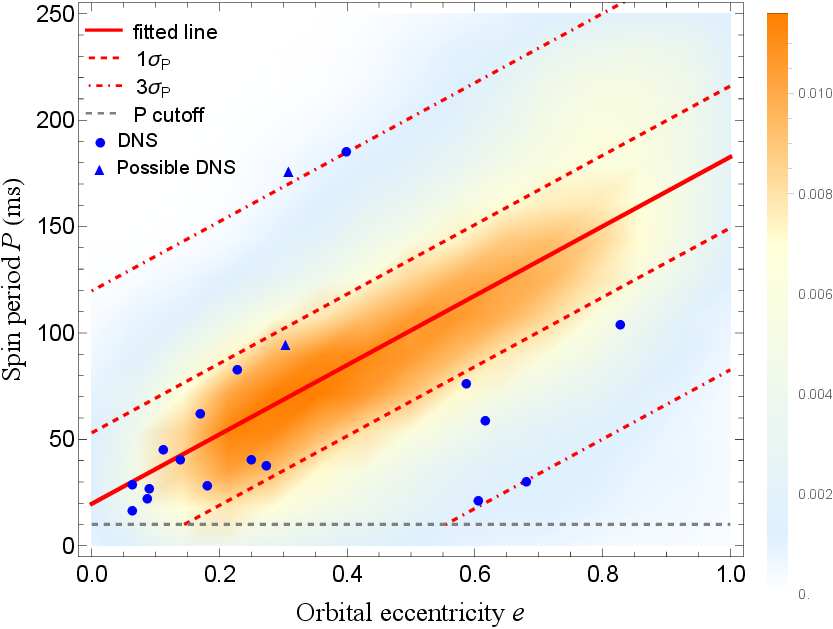}
\caption{The $P$-$e$ density distribution of simulated DNS systems, shown in a
color map, is sourced from Ref.~\cite{Dewi:2005xh}. We fit a linear model (red
solid line) to the selected data points described in Sec.~\ref{subsec-spin},
yielding an equation $P/[{\rm ms}] \simeq 163~e+19.7$, with a coefficient of
determination $R^2=0.95$ and rms error $\sigma_{\rm fit}=11.1~\rm{ms}$. The red
dashed and dot-dashed lines represent the $1\sigma_P$ and $3\sigma_P$
uncertainty bounds, respectively, while the horizontal gray dashed line marks
the lower cutoff at $P=10~\rm{ms}$. Blue dots and triangles represent known
recycled DNS systems and DNS candidates, respectively.}
\label{fig:DensityPlotFitLine}
\end{figure}

For comparison, we also present the currently discovered DNS systems in
Fig.~\ref{fig:DensityPlotFitLine}. The linear relation with added noise
effectively reproduces the simulation in Ref.~\cite{Dewi:2005xh}.

\subsubsection{NS structural parameters}

% The most commonly used deformation parameter of the NS is represented by the
% fractional difference in moments of inertia. 
In the case of a nonprecessing triaxial NS where the axis of rotation coincides
with one of its principal axes of inertia, the equatorial ellipticity of the NS
is defined as $\varepsilon \equiv (I_2-I_1)/I_3$, which can be related to the
mass quadrupole moment $Q_{22}$,  via
\begin{equation}\label{eq_eqQ22I3}
\varepsilon  = \sqrt{\frac{8\pi}{15}}\frac{Q_{22}}{I_3} \,.
\end{equation}
Here, $I_3$ represents the NS moment of inertia with respect to the principal
axis that is aligned with the spin axis, with its range adopted from theoretical
predictions in Ref.~\cite{Abbott2007}
\begin{align}
1.0 \times 10^{38} \leq I_{3}/[\rm{kg~m^2}] \leq 3.0 \times 10^{38} \,.
\end{align}

Assuming the breaking strain of the crust is at the maximum value of
$\sigma_{\max}  \approx 0.1$ for normal NS matter (neutrons, protons, and
electrons) \cite{Horowitz:2009ya}, some works have discussed the maximum mass
quadrupole moment. For example, using a chemically detailed model of the crust
in Ref.~\cite{Owen:2005fn}, one gets
\begin{align}
Q_{22,\max}/[\rm{kg~m^2}] = 2.4\times 10^{32} \,,
\end{align}
for a canonical NS with a mass of $1.4~M_{\odot}$ and a radius of $10~{\rm km}$.
It gives the maximum ellipticity the crust can support, $\varepsilon_{\max}
=2.0\times 10^{-6}$, which is consistent with the value calculated by
\citet{Ushomirsky:2000ax} ($\varepsilon_{\max} =2.8\times 10^{-6}$) and by
\citet{Morales:2022wxs} ($\varepsilon_{\max} =7.4\times 10^{-6}$).  In our
simulation below, we do not take into account the larger value of
$\varepsilon_{\max} \sim  10^{-3}$ for stars with exotic solid phases
\cite{Owen:2005fn, Haskell:2007sh, Johnson-McDaniel:2012wbj}.  To explain the
additional spin-down in the spin frequency of pulsar PSR~J1023+0038
\cite{Haskell:2017ajb}, one requires a quadrupole of 
\begin{align}
Q_{22}/[\rm{kg~m^2}] = 4.4\times 10^{28} \,.
\end{align}
We consider this the minimum value of the mass quadrupole moment, corresponding
to a minimum ellipticity $\varepsilon \approx 5.0 \times 10^{-10}$. This value
is a little smaller than the discussion on the minimum ellipticity by
\citet{Woan:2018tey}, who presented theoretical calculations of the smallest
possible mountain sizes and provided population-based evidence indicating that
millisecond pulsars exhibit a minimum ellipticity of $\varepsilon_{\min} \approx
10^{-9}$.

In the case of a precessing triaxial NS, the oblateness or poloidal ellipticity
of the NS is defined as in Eq.~(\ref{eq_smallparas}).  Since $I_1<I_2<I_3$, thus
$\varepsilon<\epsilon$. We assume that $\epsilon$ follows a range:
\begin{align}
10^{-9} \leq \epsilon \leq 10^{-5} \,.
\end{align}

The nonsphericity of the NS can be reexpressed as 
\begin{align}
\kappa = \frac{1}{16} \frac{\varepsilon}{(1-\epsilon)(\epsilon-\varepsilon)} \,,
\end{align}
which depends on the distributions of $\varepsilon$ and $\epsilon$.

The wobble angle of the NS is assumed to follow Ref.~\cite{vanEysden:2018kbc},
\begin{align}\label{eq_gammarange}
10^{-3} \leq \gamma \leq 0.05 \,.
\end{align}
The upper limit of the wobble angle corresponds to the strongest precession
candidate PSR B1828$-$11, with a wobble angle of $3^\circ$ \cite{Stairs:2000zz}.

\subsubsection{Geometry angles}

The other parameters, such as binary inclination $\iota$ (the angle between the
total angular momentum and line of sight) and opening angle of spin precession
cone $\theta_S$ (the angle between the spin and the total angular momentum)
follow the distributions below,
\begin{align}
\cos{\iota} \sim \mathcal{U}[-1, 1]  \,, \\
\cos{\theta_S} \sim \mathcal{U}[-1, 1] \,.
\end{align}
\begin{figure*}[ht]%[!htbp]
\centering
\includegraphics[scale=0.66]{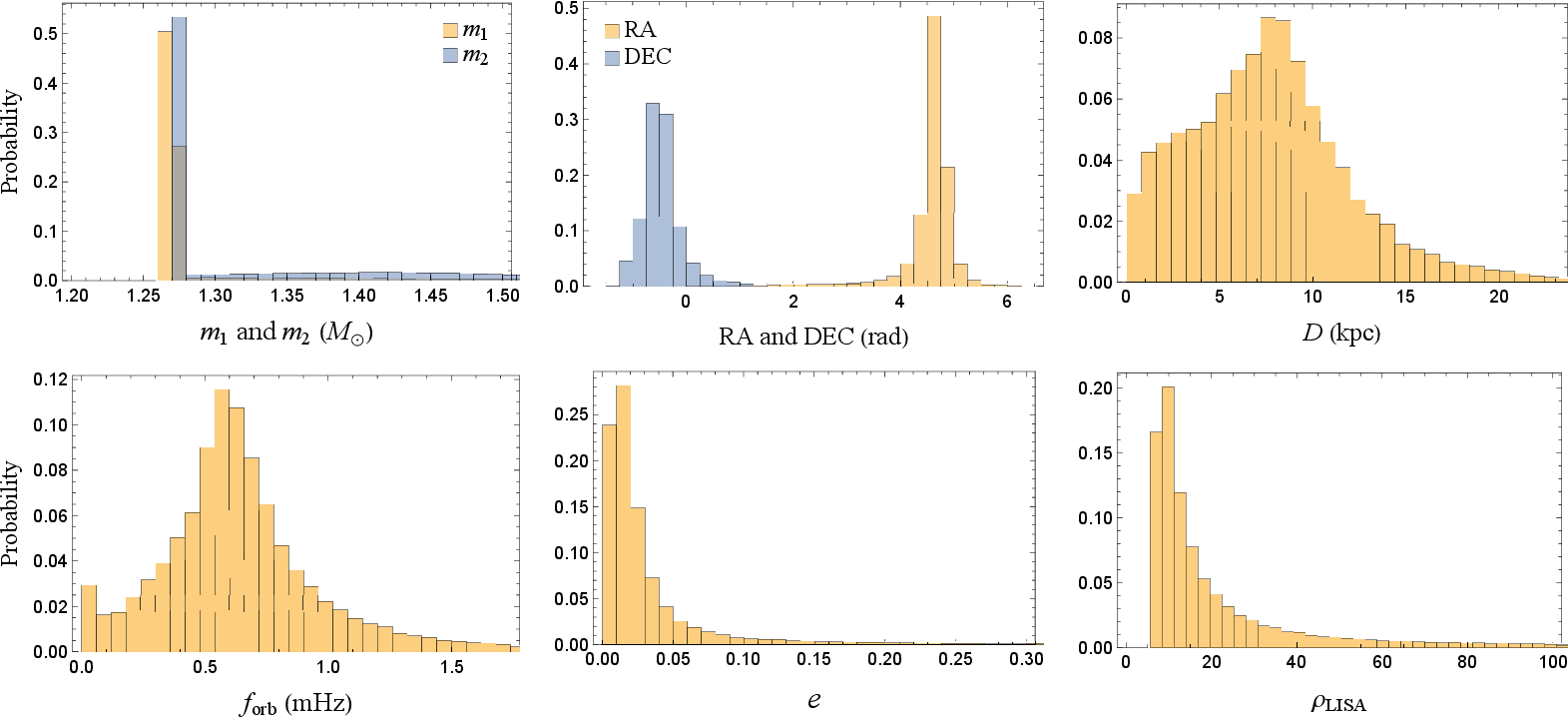}
\caption{Parameter distributions of detectable DNSs for a 4-year LISA
observation for a binary evolution model with an efficiency parameter in common
envelope phase $\alpha_{\rm ce}=10.0$ (J model) in Ref.~\cite{Wagg:2021cst}. The
properties of these systems are summarized in
Sec.~\ref{sec-DNSPopulationSynthesis}.}
\label{fig:DNSParaDistLISAalpha10}
\end{figure*}

\subsection{Population synthesis of spinning NS components in LISA-detectable DNSs}
\label{sec-NSPopulationSynthesis}

Based on population synthesis simulations of DNSs with LISA, we determine that
the optimal number of detectable DNSs within a 4-year observation period is
$\sim 35$ \cite{Wagg:2021cst}. We then generate a DNS population by sampling the
parameter distribution of these detectable sources with equal sample sizes. 
A strong correlation (Kendall $\tau=0.73$) exists between RA and DEC in the DNS population, prompting joint sampling from their two-dimensional distribution. While moderate to weak correlations are observed among orbital period, distance, and eccentricity, these and other parameters are sampled independently for simplicity.
For each DNS in this population, we assume the presence of a rapidly spinning NS and
generate the NS component population by sampling specified parameters.

The parameters of spinning NS components in LISA-detectable DNSs are primarily
categorized into two types:

\begin{enumerate}[(I)]
	\item {\bf Intrinsic parameters}
	\begin{itemize}
        \item Mass: Sampled from the NS component mass distribution of
        LISA-detectable DNSs (see Fig.~\ref{fig:DNSParaDistLISAalpha10}).
        \item Spin period: Derived from the correlation with orbital eccentricity
        via Eq.~(\ref{eq-fittedline}), incorporating Gaussian noise in
        Eq.~(\ref{eq-fittednoise}).
        \item NS structural parameters: Generated within assumed ranges using
        uniform and log-uniform sampling [see
        Eqs.~(\ref{eq_eqQ22I3}--\ref{eq_gammarange}) and Table
        \ref{tab:parameters}].
	\end{itemize}
	\item {\bf Extrinsic parameters}
	\begin{itemize}
        \item Binary orbital parameters (e.g., orbital frequency, eccentricity) and
        spatial position: Sampled from the LISA-detectable DNS distributions (see
        Fig.~\ref{fig:DNSParaDistLISAalpha10}).
        \item Polarization angle ($\psi$): Introduced by the Cosmic Explorer
        detector’s antenna pattern function, assumed to follow a uniform
        distribution $\psi \sim \mathcal{U}[0, \pi)$.
        \item Initial phase: Set to zero.
	\end{itemize}
\end{enumerate}

The simulation steps are summarized as follows:
\begin{enumerate}[(i)]
    \item Sample 35 sets of parameters  from the distributions of
    LISA-detectable DNS systems.
    \item For each sample, generate the NS spin period from orbital eccentricity
    using the pre-fitted formula (\ref{eq-fittedline}) with added Gaussian noise
    (\ref{eq-fittednoise}); apply log-uniform and uniform sampling for NS
    structural parameters; sample angular parameters within defined ranges.
    \item Repeat Steps (i) and (ii) for 1000 iterations to ensure robust
    statistical characterization, compute SNRs, and generate the final
    distributions of detectable source counts and SNRs with the Cosmic Explorer.
\end{enumerate}

\begin{table}[htb]%[htb]
    \centering
    \caption{Characteristic parameters employing uniform and log-uniform
    sampling, adopted for NS population simulation. }
    \label{tab:parameters}
    \renewcommand{\arraystretch}{1.4}
    \begin{ruledtabular}
    \begin{tabular}{lcc}
        Parameter & Uniform & Log-Uniform  \\ \hline
        Moment of inertia ($I_3/{\rm kg~m^2}$) &  $\mathcal{U}(10^{38}, 3 \times 10^{38})$ & $10^{\mathcal{U}(38, 38+\log3)}$  \\
        Oblateness ellipticity ($\epsilon$) & $\mathcal{U}(10^{-9}, 10^{-5})$  & $10^{\mathcal{U}(-9, -5)}$  \\
        Wobble angle ($\gamma$) & $\mathcal{U}(10^{-3}, 0.05)$ & $10^{\mathcal{U}(-3, -2+\log5)}$  \\
    \end{tabular}
    \end{ruledtabular}
\end{table}

\section{Detection of dual-line GW sources}
\label{sec-results}

\subsection{GW signal in the detector frame and SNR} 

After incorporating the NS Doppler shift relative to the binary barycenter and
the detector Doppler shift relative to the Solar system barycenter into the
waveform phases, and considering the detector antenna pattern functions, the GW
strain signal from a spinning triaxial NS in a tight binary can be expressed as
a sum of different waveform components \cite{Feng:2023gpe}:
\begin{align}\label{eq_ht}
h(t) = \sum_{\mu=1,2} \sum_{x=a,b,c} F_{+}(t) H_{+}^{(\mu x)}(t) + F_{\times}(t) H_{\times}^{(\mu x)}(t)   \,.
\end{align}
Here $H_{+}^{(\mu x)}(t)$ and $H_{\times}^{(\mu x)}(t)$ denote the
Doppler-modulated waveforms of $h_{+}^{(\mu x)}(t)$ and $h_{\times}^{(\mu
x)}(t)$ in Eq.~(\ref{eq_WaveformComponents}).  The antenna pattern functions of
the ground-based GW detector, $F_{+}(t)$ and $F_{\times}(t)$, are determined by
the following parameters (as detailed in Ref.~\cite{Jaranowski:1998qm}): the RA
and DEC of the source; the polarization angle of GW ($\psi$); the detector's
orientation to the local geographical directions ($\gamma_{\rm o}$); the
interferometer arm-to-arm angle ($\zeta$); the geographical latitude of the
detector location ($\lambda$); the Earth's rotational angular frequency
($\Omega_{\rm Er}$); and the initial phase of the Earth's diurnal rotation
($\phi_{\rm r}$).

Based on the 1000 independent Monte Carlo simulations in
Sec.~\ref{sec-NSPopulationSynthesis}, we set the detection SNR threshold for
Cosmic Explorer to 7 and compute the optimal SNR to select detectable sources.
The SNR for a monochromatic signal of frequency $f$ is defined as
\cite{Feng:2024ect}:
\begin{align}
\rho \equiv (h, h)^{1/2} \simeq \left[\frac{2}{S_{n}\left(f\right)} \int_{0}^{T_{\rm obs}} h(t)^2 dt \right]^{1/2} \,,
\end{align}
with $S_n(f)$ being the noise power spectral density and $T_{\rm obs}$ the
observation time.  Assuming a 40 km arm length optimized for the low-frequency
band \cite{CE2022slt}, the noise spectral density of Cosmic Explorer's
instrumental noise depends on (multiplicity of) the NS spin frequency, i.e., the
spin period addressed in Sec.~\ref{subsec-spin}. $S_n$ is evaluated at $f$ for
waveforms $h^{(1x)}$ and at $2f$ for waveforms $h^{(2x)}$. The observation time
is set to $T_{\rm obs}=4~{\rm yr}$. Angular parameters are chosen as
$\zeta=\pi/2$, $\lambda=0.764$, $\gamma_{\rm o}=1.5$, and $\phi_{\rm
r}=\phi_{\rm o}=0$ \cite{Feng:2024ect}.

\subsection{Detectable counts and SNRs with Cosmic Explorer}

\begin{figure*}[htb]%[!htbp]
\centering
\includegraphics[scale=0.7]{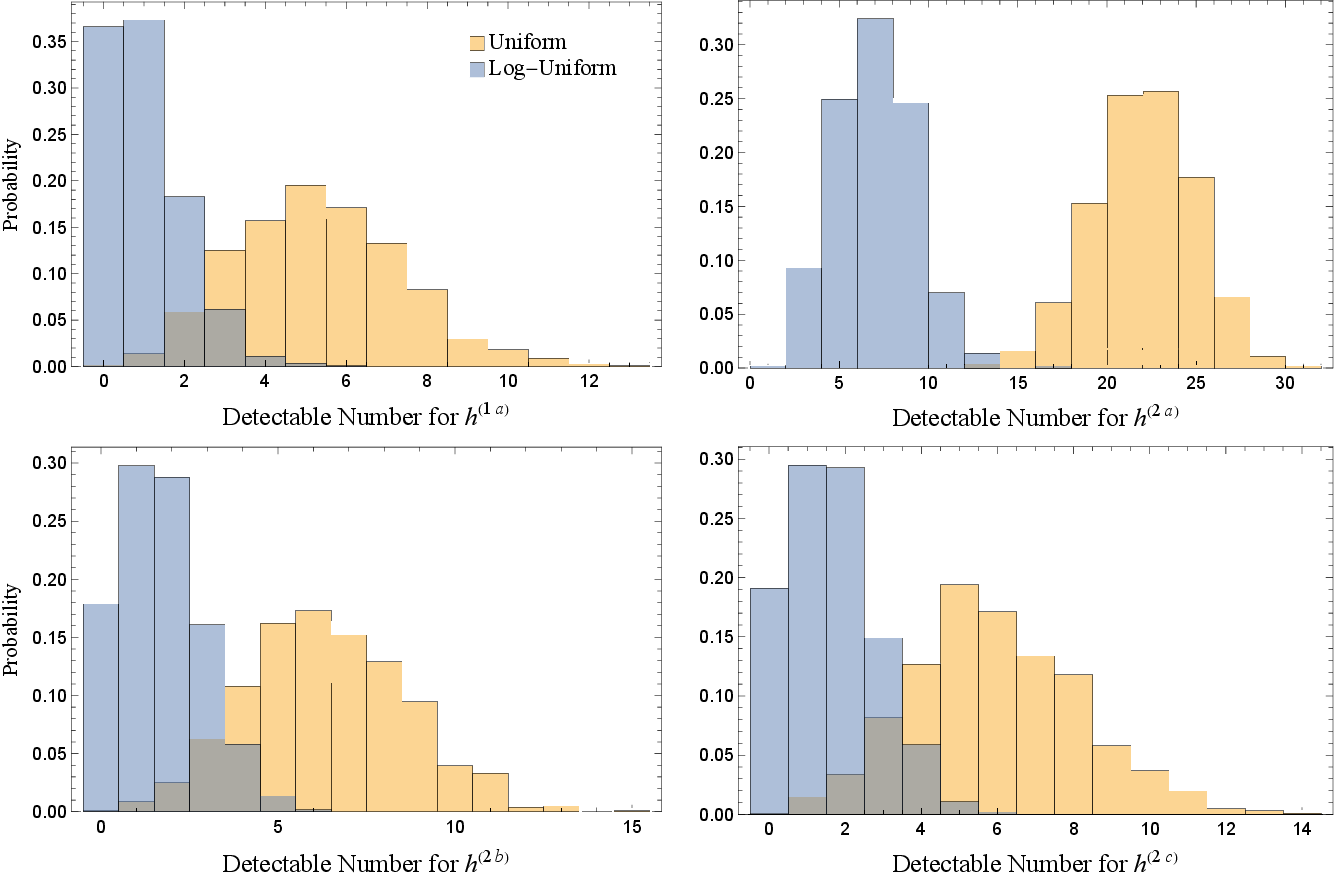}
\caption{Histograms of the number of detectable dual-line GW sources for
different waveform components, as observed by LISA and Cosmic Explorer. Blue and
orange denote log-uniform and uniform sampling of NS characteristic parameters
in the population simulation, respectively. For each panel, 1000 Monte Carlo
simulations are performed. The components $h^{(1b)}$ and $h^{(1c)}$ are excluded
from the analysis since their signal strengths fall below the detectable
threshold.  }
\label{fig:SNRCountHist}
\end{figure*}

\begin{figure*}[htb]%[!htbp]
\centering
\includegraphics[scale=0.7]{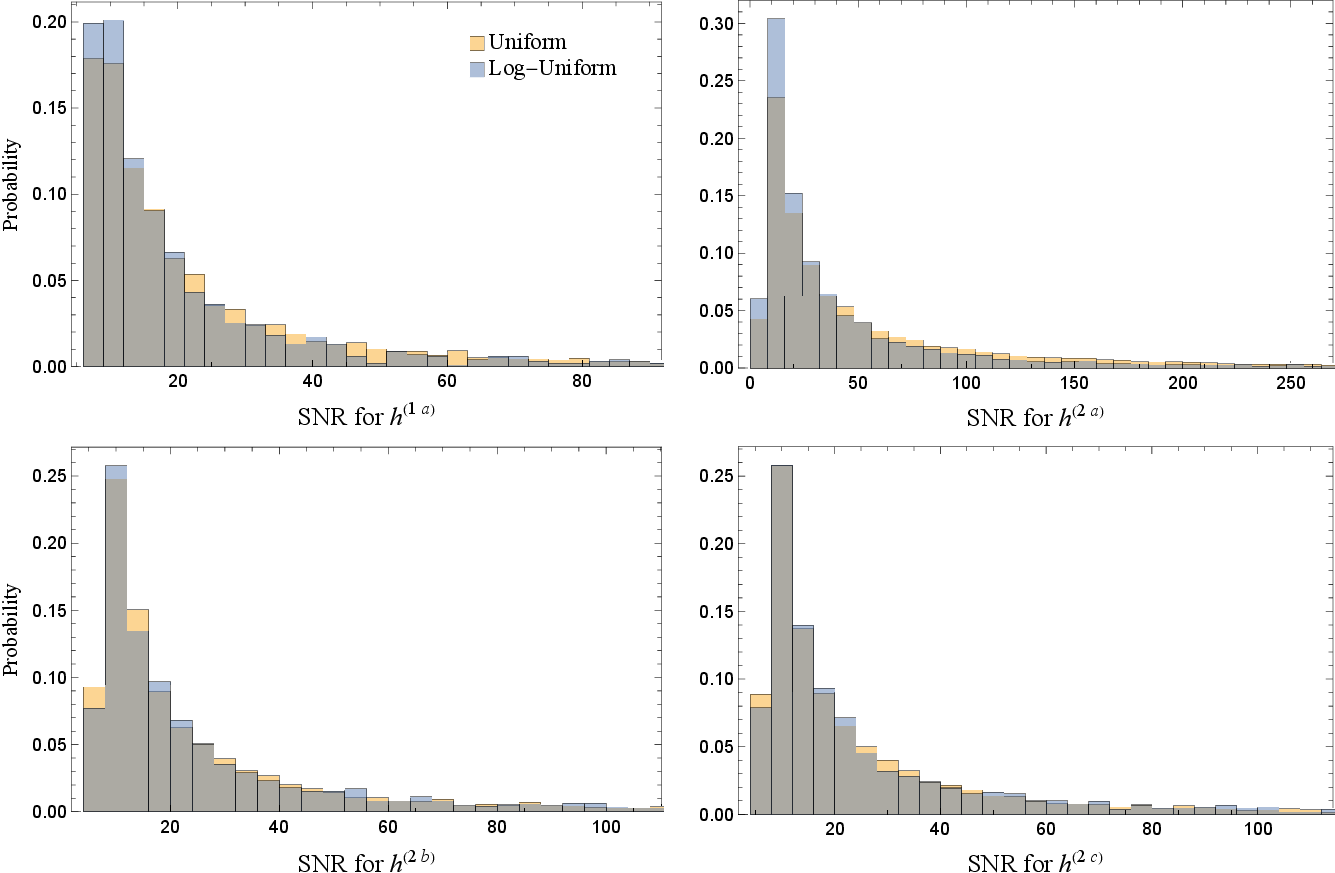}
\caption{Similar to Fig.~\ref{fig:SNRCountHist}, but showing the SNR of
detectable dual-line GW sources.}
\label{fig:SNRDistHist}
\end{figure*}

We derive the detectable counts and SNRs of various waveform components from the spinning NS population with Cosmic Explorer. 
In this study, we exclude cases where the combined SNR of all waveform components exceeds 7, as under this strategy, only one waveform component or no individual component may be detectable. Thus, the potential for inferring structural parameters via joint multi-component detection of dual-line sources in Sec.~\ref{sub_application} can be significantly reduced.
The histograms of counts and SNRs from 1000 Monte Carlo
simulations are shown in Fig.~\ref{fig:SNRCountHist} and
Fig.~\ref{fig:SNRDistHist}. The corresponding medians with $16\%-84\%$ quantile
boundaries for waveform components $h^{(1a)}$, $h^{(2a)}$, $h^{(2b)}$, and
$h^{(2c)}$ are listed in Table~\ref{tab:numberandSNR}. The signals of $h^{(1b)}$
and $h^{(1c)}$ are too weak to exceed the detection thresholds under both
sampling methods, so they are omitted from the results.

\begin{table}[htb]%[htb]
    \centering
    \caption{Detectable counts and SNRs of spinning NS population detectable by
    Cosmic Explorer from 1000 Monte Carlo simulations with uniform and
    log-uniform sampling. Values are reported as medians with $16\%$ and $84\%$
    quantile boundaries. In each parenthesis, the values from left to right are
    for waveform components $h^{(1a)}$, $h^{(2a)}$, $h^{(2b)}$, and $h^{(2c)}$.
    The signals of the $h^{(1b)}$ and $h^{(1c)}$ waveform components are too
    weak to be detected, so they are not presented here.}
    \label{tab:numberandSNR}
    \renewcommand{\arraystretch}{1.4}
    \begin{ruledtabular}
    \begin{tabular}{lcc}
    Parameter & Uniform & Log-Uniform  \\ \hline
    Number & $\left(5_{-2}^{+2}, 22_{-3}^{+3}, 6_{-2}^{+3}, 6_{-2}^{+2}\right)$  & $\left(1_{-1}^{+1}, 6_{-2}^{+3}, 2_{-2}^{+1}, 2_{-2}^{+1}\right)$  \\
    SNR    & $\left(16_{-7}^{+29}, 32_{-20}^{+102}, 16_{-7}^{+35}, 17_{-8}^{+36}\right)$  & $\left(14_{-6}^{+25}, 23_{-13}^{+64}, 17_{-8}^{+39}, 17_{-8}^{+41}\right)$  \\
    % $\Delta I_3/I_3$ & $0.08_{-0.04}^{+0.05}$  & $0.08_{-0.05}^{+0.05}$  \\
    \end{tabular}
    \end{ruledtabular}
\end{table}

The detectable counts of the spinning NS population exhibit significant
differences between two sampling methods. For the dominant waveform component
$h^{(2a)}$, uniform sampling yields a median count of 22 (in $16\%-84\%$
quantiles: 19--25), starkly higher than the 6 (in $16\%-84\%$ quantiles: 4--9)
under log-uniform sampling. This suggests that uniform sampling more effectively
captures high-amplitude components by evenly exploring parameter spaces. For
weaker components $h^{(1a)}$, the count drops from 5 (in $16\%-84\%$ quantiles:
3--7) to 1 (in $16\%-84\%$ quantiles: 0--2), and for $h^{(2b)}$ and $h^{(2c)}$,
the results show moderate counts (6 and 6 for uniform sampling; 2 and 2 for
log-uniform sampling), reflecting their intermediate sensitivity to sampling
priors. Log-uniform sampling systematically suppresses counts, likely due to its
choice toward lower parameter values, whereas uniform sampling favors broader
detectability across components.

SNR distributions align with detectability trends but exhibit nuanced
differences. The $h^{(2a)}$ component achieves a median SNR of 32 (in
$16\%-84\%$ quantiles: 12--134) under uniform sampling, surpassing the 23 (in
$16\%-84\%$ quantiles: 10--87) under log-uniform sampling, indicating stronger
signals are more efficiently captured when uniform priors are assumed. For
$h^{(1a)}$, SNRs are comparable (16 versus 14), while $h^{(2b)}$ and $h^{(2c)}$
show minor variations (16 versus 17), suggesting consistent signal behavior
across detected components. Notably, log-uniform sampling produces tighter SNR
ranges for $h^{(1a)}$ and $h^{(2a)}$, possibly due to its concentration in the
lower-amplitude parameter regions. Overall, uniform sampling gives larger SNR
for dominant signals, while log-uniform sampling gives smaller detectability.
%  yet may offer insights into rare or low-amplitude phenomena

\subsection{Application of dual-line GW detection with LISA and Cosmic Explorer}\label{sub_application}

Detections of DNSs by LISA facilitate directed searches for GWs emitted by their
spinning NS components. The next-generation, higher-sensitivity ground-based
detector Cosmic Explorer will further enhance the prospects of detecting
continuous GWs from these NSs \cite{Feng:2024ulg}.

The amplitude factors of the three waveform components $h^{(2x)}$ [cf.
Eq.~(\ref{eq_WaveformComponents})] oscillating at nearly twice the NS spin
frequency are defined as
\begin{subequations}
\begin{align}\label{eq_h20amp}
h_{2a0}&=\frac{16G b^{2}I_{3}\epsilon(\kappa - 16\kappa^{2})}{c^{4}D} \,, \\
h_{2b0}&=\frac{Gb^{2}I_{3}\epsilon(\gamma^{2}+64\kappa^{2})}{c^{4}D} \,, \\
h_{2c0}&=\frac{64G b^{2}I_{3}\epsilon\kappa^{2}}{c^{4}D} \,.
\end{align}
\end{subequations}
Upon detection of these components, the NS structural parameters can be derived
from their amplitudes via:
\begin{subequations}
\begin{align}
\gamma &= \frac{2\sqrt{(h_{2b0}-h_{2c0})h_{2c0}}}{h_{2a0}+4h_{2c0}} \,, \\
\kappa &= \frac{h_{2c0}}{4(h_{2a0}+4h_{2c0})} \,.
\end{align}
\end{subequations}

The relative measurement errors in waveform amplitudes are typically inversely
proportional to the detection SNR, i.e., 
\begin{subequations}
\begin{align}
	\Delta h_{2a0}/h_{2a0} &=1/\rho_{2a} \,, \\
	\Delta h_{2b0}/h_{2b0} &=1/\rho_{2b} \,, \\
	\Delta h_{2c0}/h_{2c0} &=1/\rho_{2c} \,.
\end{align}
\end{subequations}
Consequently, the relative errors for $\gamma$ and $\kappa$ can be derived using
the standard uncertainty propagation formula.  For a triaxially precessing NS,
the spectrum analysis of several years of observational data in principle allows
us to determine $\epsilon$ to an accuracy of $\ll 10^{-4}$ using the relation
$\Omega_{\rm p} \simeq \epsilon \Omega_{\rm r}$ (cf. Eq.~(5.7) in
Ref.~\cite{VanDenBroeck:2004wj}). This also implies that $b \simeq \Omega_{\rm
r} \cos{\gamma}$ can be measured with an accuracy of $\Delta
\cos{\gamma}/\cos{\gamma} \simeq (\Delta \gamma /\gamma){\gamma}^2 < 10^{-2}$.
Parameter estimation for DNSs with LISA or TianQin shows that the relative error
in the source distance, derived from chirp mass, orbital period, and amplitude
factor, is approximately the inverse of the SNR (see Eqs.~(31) and (33) in
Ref.~\cite{Feng:2023fez}), ${\Delta D}/{D} ={1}/{\rho_{\rm DNS}}$.  For the more
detectable waveform components $h^{(2a)}$(characterized by higher SNR), applying
the standard uncertainty propagation to $h_{2a0}$ in Eq.~(\ref{eq_h20amp})
yields the relative measurement accuracy for the moment of inertia:
\begin{align}
\frac{\Delta I_{3}}{I_{3}} = \sqrt{\frac{1}{\rho_{2a}^{2}} + (1 - 32\kappa)^{2}\left(\frac{1}{\rho_{2a}^{2}} + \frac{1}{\rho_{2c}^{2}}\right) + \frac{1}{\rho_{\rm DNS}^{2}}} \,.
\end{align}

\begin{figure}[t]%[!htbp]
\centering
\includegraphics[scale=0.66]{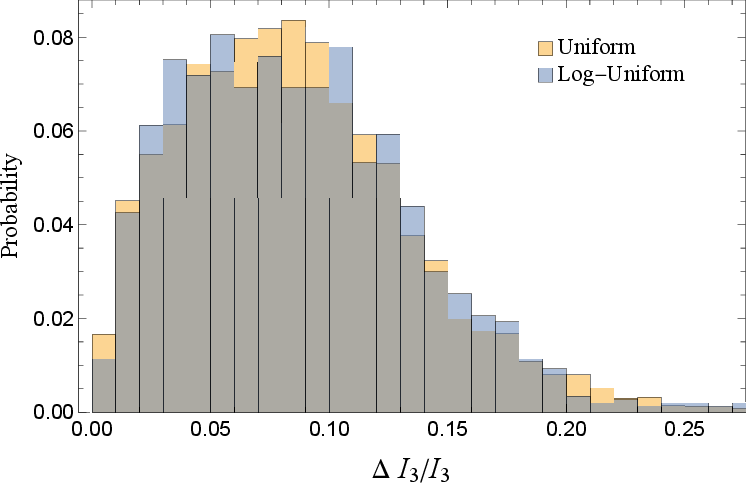}
\caption{The relative measurement accuracy of the moment of inertia is inferred
from the joint detection of dual-line GW sources using LISA and Cosmic Explorer.
For each sampling, 1000 Monte Carlo simulations are performed. The estimated
distributions show a similar trend under uniform sampling (orange) and
log-uniform sampling (blue) for the NS population simulation.}
\label{fig:DeltaI3toI3Plot}
\end{figure}

As shown in Fig.~\ref{fig:DeltaI3toI3Plot}, the relative measurement accuracy of
the moment of inertia---derived from joint detections of dual-line GW sources
with LISA and Cosmic Explorer---is evaluated via 1000 Monte Carlo simulations.
The results indicate a median relative accuracy of $8_{-4}^{+5}\%$ (with
$16\%-84\%$ quantiles) in both sampling methods, demonstrating the insensitivity
to the sampling strategy for this NS characteristic parameter.

\section{Conclusions}\label{sec-conclusion}

In this study we investigate the detectability and scientific applications of Galactic DNS systems as dual-line GW sources. First, we expand the waveforms from spinning NSs undergoing spin-orbit coupling up to $\mathcal{O}(\gamma^2, \kappa^2)$ in the small NS structural parameters $\gamma$ and $\kappa$. We then synthesize the population of spinning NS components with LISA-detectable DNS systems, estimate their detectable counts and SNRs, and finally explore their applications for measuring the moment of inertia of NSs.

In simulating the NS component population in these systems, we employ two sampling methods for NS characteristic parameters: uniform and log-uniform. The results reveal a clear trend: uniform sampling outperforms log-uniform sampling in detecting high-amplitude components and achieving higher SNRs,  due to its assumption for exploration of the parameter space. While log-uniform sampling reduces detectability, it may offer insights into rare or low-amplitude phenomena when prior knowledge suggests distributions with smaller parameters. These findings emphasize the importance of prior knowledge about NS structural parameters in NS population synthesis for GW detection.
Once dual-line DNSs are detected, we can use the Bayes factor to estimate which parameter distribution is preferred. The Bayes factor quantifies the relative evidence of data for two different models, and it accumulates multiplicatively with the detection number, thereby strengthening evidence for one model.

Although we neglected the spin-down effects due to gravitational and electromagnetic radiation in our waveform model [cf. Eq.~(\ref{eq_WaveformComponents})], based on estimates by \citet{Feng:2024ect}, we can still add a frequency derivative term to our waveforms following the parameterization currently employed in continuous wave searches for NSs. Given the negligible impact of this term on the radiation frequency, we anticipate that it will not significantly affect the estimation distributions presented in this work.

It should be noted that the results of our dual-line GW simulations are contingent on the simulated outcomes for LISA-detectable DNSs. While LISA may detect $\sim 35$ DNSs in the optimistic case, distinguishing them from numerous double WD systems remains challenging—--\citet{Wagg:2021cst} suggests only $\sim 60\%$ of DNSs can be reliably identified due to parameter overlaps. 
Under a more conservative model assumption (e.g., 8 detectable LISA DNS systems over a 4-year observation, as reported in Figure 8 of \cite{Wagg:2021cst}), the predicted count of dual-line sources would decrease to roughly one-fourth of the current optimistic estimate.
Additionally, binary population simulations assume DNSs remain at their birth locations, neglecting supernova kick effects on positions; however, small kicks from electron-capture supernovae mitigate this impact. Future studies using Cogsworth \cite{Wagg_2025} (combining population synthesis and galactic dynamics) will refine these assumptions.

For future investigation, a more self-consistent binary star simulation framework could utilize POSYDON \cite{Fragos:2022jik, Andrews:2024saw}, a novel binary population synthesis code that incorporates the stellar evolution models of MESA \cite{Paxton:2015jva, Paxton:2019lxx}. Additionally, COMPAS has recently advanced to incorporate NS spin evolution into its codebase, following the theoretical modeling by \citet{Chattopadhyay:2019xye}. These software tools enable the simultaneous simulation of binary orbital and stellar spin evolution, facilitating more detailed assessments of detectability for both space-borne and ground-based GW detectors.

\begin{acknowledgments}

We gratefully acknowledge Tom Wagg for providing the LISA-detectable DNS population simulation data used in this study and for his careful review of the manuscript, which enhanced its quality through insightful comments.
We thank Yan Wang, Tan Liu, Hongbo Li, Ziming Wang, Ze Zhang, Yong Shao, and En-Kun Li for helpful discussions on NS population simulation.    
This work is supported by the
National Natural Science Foundation of China (12447109), the Beijing Natural
Science Foundation (1242018), the National SKA Program of China
(2020SKA0120300), the Max Planck Partner Group Program funded by the Max Planck
Society, and the High-Performance Computing Platform of Peking University.

\end{acknowledgments}

\appendix

% \section{Fitting of the spin period–orbital eccentricity relation of DNS systems}
% \label{App:pefitting}

\bibliography{reference}

\end{document}